\documentclass[12pt]{article}
\usepackage{graphicx}
\usepackage{epstopdf}
\usepackage{caption}
\usepackage{subcaption}
\usepackage{subfig}

\usepackage{epsfig,amsmath,amssymb,verbatim,mathrsfs}

\def\beq{\begin{eqnarray}}
\def\eeq{\end{eqnarray}}
\def\bea{\begin{eqnarray}}
\def\eea{\end{eqnarray}}
\newcommand{\s}{\text{\tiny S}}
\newcommand{\sm}{\text{\tiny SM}}

\newcommand{\dm}{\text{\tiny DM}}
\newcommand{\f}{\mathcal{F}}

\renewcommand{\thefootnote}{\roman{footnote}}

\begin{document}

\setlength{\baselineskip}{0.2in}


\begin{titlepage}
\noindent
\vspace{0.2cm}

\begin{center}
  \begin{Large}
    \begin{bf}
A Class of Three-Loop  Models with Neutrino Mass and Dark Matter

     \end{bf}
  \end{Large}
\end{center}

\vspace{0.2cm}

\begin{center}

\begin{bf}

{Chian-Shu~Chen,$^{1,2,}$\footnote{chianshu@phys.sinica.edu.tw}
Kristian~L. McDonald$^{3,}$\footnote{klmcd@physics.usyd.edu.au} and
Salah~Nasri$^{4,5,}$\footnote{snasri@uaeu.ac.ae}}\\
\end{bf}
\vspace{0.5cm}
 \begin{it}
$^1$ Physics Division, National Center for Theoretical Sciences, Hsinchu, Taiwan 300\\
\vspace{0.1cm}
$^2$ Department of Physics, National Tsing Hua University, Hinschu, Taiwan 300\\
\vspace{0.1cm}
$^3$ ARC Centre of Excellence for Particle Physics at the Terascale,\\
School of Physics, The University of Sydney, NSW 2006, Australia\\
\vspace{0.1cm}
$^4$ Physics Department, UAE University, POB 17551, Al Ain, United Arab Emirates\\
\vspace{0.1cm} 
$^5$ Laboratoire de Physique Theorique, ES-SENIA
University, DZ-31000 Oran, Algeria \vspace{0.3cm}
\end{it}
\vspace{0.5cm}

\end{center}


\begin{abstract}

We study a class of three-loop models for neutrino mass in which  dark matter plays a key role in enabling the mass diagram. The simplest models in this class have Majorana dark matter and include the proposal of Krauss, Nasri and Trodden; we identify the remaining related models, including the viable colored variants. The next-to-simplest models use either more multiplets and/or a slight modification of the loop-diagram, and predict inert N-tuplet scalar dark matter.

\end{abstract}

\vspace{1cm}

\end{titlepage}

\renewcommand{\thefootnote}{\arabic{footnote}}
\setcounter{footnote}{0} %
\setcounter{page}{1} 


\vfill\eject


\section{Introduction\label{sec:introduction}}

In recent years  the idea that the origin of neutrino mass and the existence of dark matter (DM) may be related has received much attention. The neutrino mass and DM problems are perhaps our most compelling pieces of evidence for physics beyond the Standard Model (SM), and it is therefore reasonable to consider unified solutions to these problems.

A simple model predicting a connection between these issues was proposed by Ma~\cite{Ma:2006km}. This model achieves one-loop neutrino mass with the DM being either an inert scalar-doublet or a Majorana fermion. The model is well studied in the literature~\cite{Ho:2013hia}. In particular, it was shown that the model belongs to a larger class of models, all of which achieve neutrino mass by a loop-diagram with the same topology, while also giving DM candidates~\cite{Law:2013saa}. One of the  related models  uses a Majorana triplet-fermion~\cite{Ma:2008cu}, while the others employ Dirac fermions~\cite{Law:2013saa}. The latter models must have scalar DM, with singlet, doublet and triplet cases possible~\cite{Law:2013saa}.

An earlier model proposing  a  common solution to the DM and neutrino mass problems was advocated by Krauss, Nasri and Trodden (KNT)~\cite{Krauss:2002px} (for detailed studies see Refs.~\cite{Baltz:2002we,Cheung:2004xm,Ahriche:2013zwa,Ahriche:2014xra}). This model achieves neutrino mass at the three-loop level and predicts Majorana singlet-fermion DM. In analogy with the one-loop models, it is natural to ask if the KNT model could also belong to a larger class of three-loop models with DM candidates. In this paper we perform a systematic study for variants of the KNT model. We first consider generalizations that employ Majorana fermions, identifying the viable models and, in particular, presenting the viable colored-variants. We then show that Dirac fermions can also be used  to generate a radiative mass-diagram with the same topology. The latter models require inert scalar DM, different from the KNT (and related) models.

The plan of this paper is as follows. In Section~\ref{sec:KNTmodel} we briefly summarize the KNT model and discuss related models in the literature. A systematic classification of the minimal variants of the KNT model is performed in Section~\ref{sec:3loop_withDM}. Section~\ref{sec:distinct_SM_fermion} discovers variants with Dirac mediators that achieve neutrino mass by a loop diagram with the same topology.  Modifying the loop diagram slightly, we show that additional variants are possible in Section~\ref{sec:general_3loop_withDM}. We conclude in Section~\ref{sec:conclusion}. Before proceeding we note that a number of other works have studied models with connections between neutrino mass and DM; for a selection see Refs.~\cite{Aoki:2008av,Aoki:2011yk,Kajiyama:2013zla,Kanemura:2011mw}. Also, there may  be other interesting three-loop topologies beyond those considered here, in line with the general treatment of effective operators with $\Delta L=2$~\cite{Angel:2012ug}. For a general discussion of neutrino mass see Ref.~\cite{Boucenna:2014zba}.

\section{The KNT Model\label{sec:KNTmodel}}

KNT  proposed a simple model with a connection between the existence of massive neutrinos and DM~\cite{Krauss:2002px}. The SM is extended to include the exotic scalars $S\sim(1,1,2)$ and $\phi\sim(1,1,2)$, and the fermions $\f_{iR}\sim(1,1,0)$, where $i$ labels fermion generations.  A discrete ($Z_2$) symmetry is also imposed, such that $\phi$ and $\f$ are $Z_2$-odd, $\{\phi,\,\f\}\rightarrow\{-\phi,\,-\f\}$,  while $S$  and the SM fields are $Z_2$-even. The Lagrangian then includes the terms
\bea
\mathcal{L}&\supset& \mathcal{L}_{\sm}\,+\{f_{\alpha\beta}\,\overline{L^c_\alpha}\, L_\beta\, S^++g_{i\alpha}\,\overline{\f^c_{i}} \,\phi^+ \,e_{\alpha R}+\mathrm{H.c}\}\;-\;\frac{1}{2}\,\overline{\f_i^c} \,\mathcal{M}_{ij}\, \f_j\;-\;V(H,S,\phi),\label{eq:KNT_Lagrangian}
\eea
where the mass matrix is taken diagonal, without loss of generality; $\mathcal{M}=\mathrm{diag}(M_1,\,M_2,\,M_3)$. We order the masses as $M_1<M_2<M_3$, and use Greek letters to label SM flavors, $\alpha,\,\beta\in\{e,\,\mu,\,\tau\}$. 

\begin{figure}[ttt]
\begin{center}
        \includegraphics[width = 0.6\textwidth]{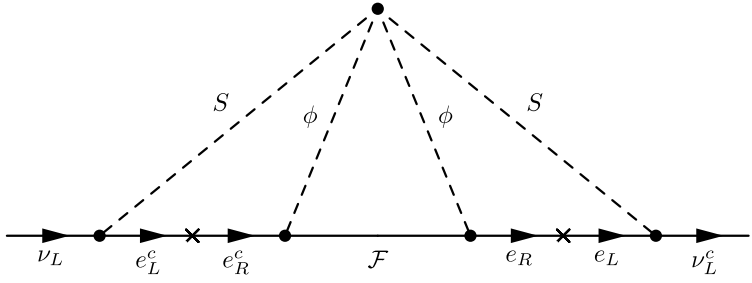}
\end{center}
\caption{A three-loop diagram for radiative neutrino mass, where $S$ and $\phi$ are beyond-SM scalars and $\mathcal{F}$ is a beyond-SM fermion.}\label{fig:KNT_nuDM5}
\end{figure}

The scalar potential contains the terms
\bea
V(H,\,S,\,\phi)&\supset&\frac{\lambda_\s}{4} (S^*)^2\phi^2+\mathrm{H.c.},\label{eq:KNT_pot_Lbreaking}
\eea
and the combination of Eqs.~\eqref{eq:KNT_Lagrangian} and \eqref{eq:KNT_pot_Lbreaking} explicitly breaks lepton number symmetry. This results in Majorana neutrino masses at the three-loop level, as shown in Figure~\ref{fig:KNT_nuDM5}. Calculating the loop diagram gives the mass matrix as
\bea
(\mathcal{M}_\nu)_{\alpha\beta}&=& \frac{\lambda_\s}{(4\pi^2)^3}\,\frac{m_\sigma m_\rho}{M_\phi}\,f_{\alpha\sigma}\,f_{\beta\rho}\,g^*_{\sigma i}\,g^*_{\rho i}\times F\left(\frac{M_i^2}{M_\phi^2},\frac{M_\s^2}{M_{\phi}^2}\right),
\eea
where $F(x,y)$ is a function that encodes the loop integrals, whose explicit form is given in Ref.~\cite{Ahriche:2013zwa}.

The $Z_2$ symmetry plays two roles in the model. Firstly, it prevents the Yukawa term $\bar{L}\tilde{H}F^c_{L}$, which would otherwise produce tree-level neutrino masses via a (Type-I) seesaw mechanism. Secondly, the lightest $Z_2$-odd field is absolutely stable. Provided this is the lightest neutrino $\f_1$, the model contains a viable DM candidate~\cite{Ahriche:2013zwa} and gives a unified solution to the DM and neutrino-mass problems. The DM and $Z_2$-odd fields must be relatively light, with $M_1<225$~GeV and $M_\phi<245$~GeV, while the combination of neutrino experiments and the DM relic-density prefers $M_\s>M_\phi$. The model can be probed at collider experiments~\cite{Ahriche:2014xra} and can modify the branching fraction for Higgs decays to $2\gamma$ and $Z\gamma$. The signal from flavor-changing decays such as $\mu\rightarrow e+\gamma$ may be observable in future experiments~\cite{Ahriche:2013zwa}. In this model  the DM is sequestered from SM neutrinos and propagates in the inner loop of the mass diagram.

\subsection{Triplet Variant of the KNT Model\label{subsec:triplet_KNT}}

The seesaw mechanism can be generalized to a triplet (or Type-III) variant that employs $SU(2)_L$ triplet fermions with vanishing hypercharge~\cite{Foot:1988aq}. Similarly, it was recently shown that the KNT model can be generalized to a triplet variant~\cite{Ahriche:2014cda}. One retains the scalar $S$ but $\phi$ and $\f$ are now $SU(2)_L$ triplets, $\phi\sim(1,3,2)$ and $\f\sim(1,3,0)$. The $Z_2$ symmetry  is retained, $\{\phi,\,\f\}\rightarrow\{-\phi,\,-\f\}$, with all other fields being $Z_2$-even. The Lagrangian again contains the terms in Eq.~\eqref{eq:KNT_Lagrangian}, with $\f_{i}$ as triplet fermions, and the potential contains terms similar to \eqref{eq:KNT_pot_Lbreaking},
\bea
V(H,S,\phi)&\supset& \frac{\lambda _{\text{{\tiny S}}}}{4}(S^{\ast})^{2}\phi_{ab}\phi_{cd}\epsilon^{ac}\epsilon^{bd}+\frac{\lambda _{\text{{\tiny S}}}^{\ast}}{4}(S)^{2}(\phi^{\ast})^{ab}(\phi^{\ast })^{cd}\epsilon _{ac}\epsilon _{bd}.\label{eq:triplet_KNT_pot_Lbreaking}
\eea
We write the triplet as a symmetric matrix,
\bea
\phi_{11}=\phi^{++},\quad \phi_{12}=\phi_{21}=\frac{1}{\sqrt{2}}\,\phi^{+},\quad
\phi_{22}=\phi^{0}.
\eea
The combination of these terms again breaks lepton-number symmetry, giving radiative neutrino mass at the three-loop level. The Feynman diagram has the same form as Figure~\ref{fig:KNT_nuDM5}, except now there are three distinct diagrams with different sets of triplet fields propagating in the inner loop~\cite{Ahriche:2014cda}.

In this model the $Z_2$ symmetry again prevents tree-level neutrino mass via a (Type-III) seesaw mechanism, and ensures a stable DM candidate. The DM is the lightest neutral triplet-fermion, $\f_1^0$, as $\phi^0$ DM is excluded by direct-detection experiments. The DM should have a mass $M_\dm\sim2$~TeV, making both $\phi$ and $\f$ too heavy to be probed at the LHC. However, the scalar $S$ may be sufficiently light to appear at colliders, with $M_\s=\mathcal{O}(10^2)$~GeV  found to be consistent with the demands of neutrino experiments and the DM relic-density~\cite{Ahriche:2014cda}. Flavor changing effects can also appear in next-generation experiments. The model is therefore a testable variant of the KNT proposal.

\subsection{Larger Representations\label{subsec:large_KNT}}

The seesaw mechanism can also be generalized to a quintuplet variant~\cite{Kumericki:2012bh,Liao:2010cc,McDonald:2013kca}. Similarly the KNT model can be generalized to a variant employing the fermion $\f\sim(1,5,0)$, and the scalar $\phi\sim(1,5,2)$~\cite{Ahriche:2014oda}. In these cases the most-general Lagrangian contains the terms in Eq.~\eqref{eq:KNT_Lagrangian}, as well as terms similar to Eq.~\eqref{eq:triplet_KNT_pot_Lbreaking} which break lepton number symmetry and give three-loop neutrino mass via the diagram in Figure~\ref{fig:KNT_nuDM5} (there are now five diagrams with different sets of fields in the inner loop).

There is one important difference, however, for the model with larger multiplets. Now the $Z_2$ symmetry need not be imposed to preclude tree-level neutrino masses. Thus, the quintuplet variant is a viable radiative model of neutrino-mass, irrespective of DM considerations. Also, the most-general Lagrangian contains a single $Z_2$ symmetry-breaking term, so the model contains a softly broken accidental $Z_2$ symmetry. In the limit that a single parameter vanishes, $\lambda\rightarrow0$, this symmetry becomes exact and the lightest $Z_2$-odd field is a stable DM candidate. Even for $\lambda\ne0$, one can always choose $\lambda\ll1$ to obtain long-lived DM without imposing the $Z_2$ symmetry~\cite{Ahriche:2014oda}. This feature differs from the KNT model and the triplet variant.  In the analysis that follows we  restrict our attention to multiplets no larger than the adjoint, though related generalizations may be possible if this restriction is relaxed.

\section{A Class of Three-Loop Models with Dark Matter\label{sec:3loop_withDM}}

\begin{figure}[ttt]
\begin{center}
        \includegraphics[width = 0.6\textwidth]{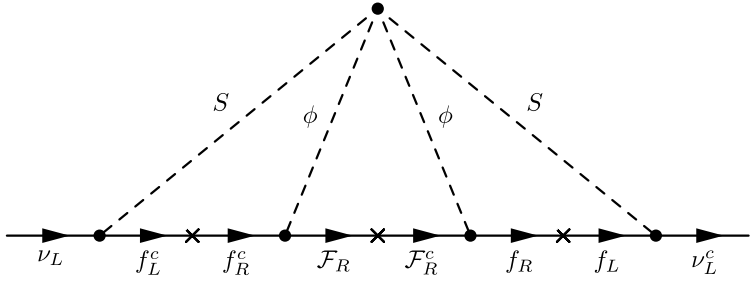}
\end{center}
\caption{A three-loop diagram for radiative neutrino mass, where $S$ and $\phi$ are beyond-SM scalars and $\mathcal{F}$ is a beyond-SM fermion. DM propagates in the inner loop.}\label{fig:general_KNT}
\end{figure}
We seek  generalizations of the KNT model that retain the following features: ($i$) The models contain $Z_2$-odd fields, including the DM, that propagate in the inner loop of the neutrino mass diagram. ($ii$) The internal fermions in the outer loops are SM fields. ($iii$) The DM is non-colored. The generalized Feynman diagram for neutrino mass in this class of models appears in Figure~\ref{fig:general_KNT}, where $\f$ and $\phi$ are $Z_2$-odd and  $S$ is $Z_2$-even. Here $f_{L,R}$ denotes a SM fermion. There are six cases to consider:
\begin{itemize}

\item $f^c_{L,R}=u^c_{L,R}$ being an up-type quark. The outer-left vertex then results from the operator $\overline{Q^c}L S_1$, where $Q$ is the SM quark doublet. In this case the diagram cannot be closed without breaking gauge invariance so neutrino mass via Figure~\ref{fig:general_KNT} is not possible.\footnote{This is contrary to the claims of Ref.~\cite{Ng:2013xja} which uses $f_{L,R}=u_{L,R}$. The resulting model possesses a lepton number symmetry under which only $L$, $e_R$ and $S$ transform ($S$ is labeled as $\chi$ in that work). The Yukawa Lagrangian is invariant under this symmetry and the potential is a function of the modulus $S^\dagger S$, so the symmetry remains unbroken in the full Lagrangian. This is sufficient to prevent the Majorana neutrino masses claimed in Ref.~\cite{Ng:2013xja}.} 

\item $f^c_{L,R}=u_{R,L}$ being an up-type quark. As with the previous case, it is not possible to successfully close the diagram while maintaining gauge invariance.

\item $f^c_{L,R}=e_{R,L}$ being a charged lepton. Then $S\sim(1,2,1)$ has the same quantum numbers as the SM scalar doublet, as does $\phi\sim(1,2,1)$. In this case the three-loop diagram can be successfully realized. However, the model always allows a one-loop diagram that is expected to dominate; for example, with $\f\sim(1,1,0)$ one also obtains the one-loop diagram from Ref.~\cite{Ma:2006km}, which is expected to dominate the three-loop diagram.  The case $f_{L,R}=e_{R,L}^c$ is therefore not viable as a three-loop model of neutrino mass.


\begin{table}
\centering
\begin{tabular}{|c|c|c|c|c|c|}\hline
\ \ Model\ \ &
$f_{L,R}$&$\mathcal{F}_R$  & $S$&$\phi$&Dark Matter \\
\hline
KNT~\cite{Krauss:2002px}&$e_{L,R}$& $\ \ (1,1,0)\ \ $ &$\ \ (1,1,2)\ \ $&\ \ $(1,1,2)$\ \ & $\f_R^0$\\ 
\hline
Triplet  Variant of KNT~\cite{Ahriche:2014cda}&$e_{L,R}$& $\ \ (1,3,0)\ \ $ &$\ \ (1,1,2)\ \ $&\ \ $(1,3,2)$\ \ &$\f_R^0$ \\ 
\hline
New Model& $d_{R,L}^c$&$(1,1,0)$ &$(3,2,1/3)$&\ $(\bar{3},2,-1/3)$\ & $\f_R^0$\\ 
\hline
New  Triplet Variant&$d_{R,L}^c$& $(1,3,0)$ &$(3,2,1/3)$&$(\bar{3},2,-1/3)$&$\f_R^0$\\ 
\hline
New Model& $d_{L,R}$&$(1,1,0)$ &$(\bar{3},1,2/3)$&\ $(3,1,-2/3)$\ & $\f_R^0$\\ 
\hline
New Color-Triplet Variant&$d_{L,R}$& $(1,1,0)$ &$(\bar{3},3,2/3)$&$(3,1,-2/3)$&$\f_R^0$\\ 
\hline
\end{tabular}
\caption{\label{KNT_variant_table} Models with radiative neutrino mass via Figure~\ref{fig:general_KNT} with DM propagating in the inner loop. Here $f$ ($\f$) is a SM (beyond-SM) fermion while  $S$ and $\phi$ are beyond-SM scalars. In all cases the DM is a neutral Majorana fermion.}
\end{table}

\item $f^c_{L,R}=d_{R,L}$ being a down-type quark. This case can generate neutrino mass with $ S\sim(3,2,1/3)$ and $\phi\sim(\bar{3},2,-1/3)$. The fermion can be a triplet or a singlet, $\f_R\sim(1,R_\f,0)$, with $R_\f=1$ or $R_\f=3$. The Lagrangian  contains the terms
\bea
\mathcal{L}&\supset& \{f_{\alpha'\alpha}\,\overline{d_{\alpha'R}}\, L_\beta\, S+g_{i\alpha'}\,\overline{\f_i}\,\phi\, Q_{\alpha'}\, +\mathrm{H.c}\}\;-\;\frac{1}{2}\,\overline{\f_i^c} \,\mathcal{M}_{ij}\, \f_j\;-\;V(H,S,\phi),\label{eq:dR_KNT_Lagrangian}
\eea
where $\alpha'$ labels quark flavors and the potential includes the terms:
\bea
V(H,S,\phi)&\supset& \lambda_s (S^\dagger \phi^\dagger)^2 +\mathrm{H.c.}
\eea
The combination of these terms breaks lepton number symmetry and generates neutrino mass at three-loops via Figure~\ref{fig:general_KNT}. The neutral field $\f_1^0$ is the only DM candidate. 

\item $f^c_{L,R}=d^c_{L,R}$ being a down-type quark. One also obtains a successful model in this case. The SM is extended to include the scalars $S\sim(\bar{3},R_S,2/3)$ and $\phi\sim(3,1,-2/3)$, where $R_S=1$ or $R_S=3$, along with the fermion $\f_R\sim(1,1,0)$.
The Lagrangian  contains the terms\footnote{Here the SM quarks are denoted by $\overline{Q^c}=-Q^T_a C^{-1} \epsilon^{ab} \sim(3,\bar{2},1/3)$ and $d_R^c=C\overline{d}_R^T\sim(\bar{3},1,4/3)$.}
\bea
\mathcal{L}&\supset& \{f_{\alpha'\alpha}\,\overline{Q^c_{\alpha'}}\, L_\beta\, S+g_{i\alpha'}\,\overline{\f_{i}}\, \phi\,d^c_{\alpha' R}+\mathrm{H.c}\}\;-\;\frac{1}{2}\,\overline{\f_i^c} \,\mathcal{M}_{ij}\, \f_j\;-\;V(H,S,\phi),\label{eq:dR_KNT_Lagrangian'}
\eea
where the potential again contains the terms:
\bea
V(H,S,\phi)&\supset& \lambda_s (S^\dagger \phi^\dagger)^2 +\mathrm{H.c.}
\eea
These terms break lepton number symmetry and give the desired three-loop diagram. The DM is the lightest neutral fermion $\f_1^0$.

\item $f_{L,R}=e_{L,R}$ gives the KNT model (or the triplet variant) when $\f$ is a Majorana fermion.
\end{itemize}

Note that in drawing Figure~\ref{fig:general_KNT} we assumed the following: Both occurrences of $\phi$ are the same multiplet; both occurrences of $S$ are the same multiplet, and; $\f$ is a Majorana fermion (vanishing hypercharge). We initially relaxed all of these assumptions but found that whenever the same SM fermion $f$ appears in both the left and right loops, as drawn in Figure~\ref{fig:general_KNT}, one cannot obtain a successful model with a Dirac fermion (nonzero hypercharge) and distinct scalar multiplets. We therefore restricted our attention to the viable case in Figure~\ref{fig:general_KNT}. We turn to variants with a Dirac fermion $\f$ in the next section.

To summarize, the only viable models with non-colored DM giving neutrino mass at the three-loop level by Figure~\ref{fig:general_KNT} are the KNT model, its triplet variant, and four new models employing either $f_{L,R}=d_{R,L}^c$ or $f_{L,R}=d_{L,R}$. These results are summarized in Table~\ref{KNT_variant_table}. All of these models predict Majorana DM,  with four cases giving singlet-fermion DM and two giving triplet-fermion DM. In the latter case we expect the DM to be $M_\dm\sim 2$~TeV, in line with previous studies of triplet-fermion DM~\cite{Ahriche:2014cda}. The precise allowed range will vary a little among models, due to the extra couplings and annihilation channels but, based on the recent analysis of the triplet-KNT model~\cite{Ahriche:2014cda}, the order-of-magnitude estimate for the DM should not significantly change (due to the sizable common contribution of $SU(2)_L$ gauge-interactions to the annihilation cross sections). The singlet-fermion DM is expected to be $M_\dm=\mathcal{O}(10^2)$~GeV, with some sensitivity to new annihilation channels in the models.

\section{Models with Dirac Mediators: Allowing Distinct SM Fermions\label{sec:distinct_SM_fermion}}

\begin{figure}[ttt]
\begin{center}
        \includegraphics[width = 0.6\textwidth]{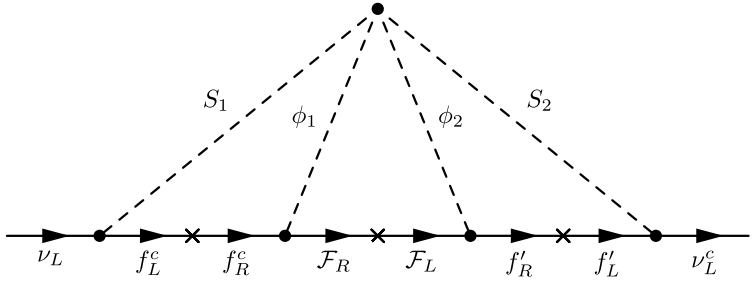}
\end{center}
\caption{A three-loop diagram for radiative neutrino mass, where $S_{1,2}$ and $\mathcal{F}$ are beyond-SM fields and DM propagates in the inner loop. The simplest case has inert-doublet DM [$S_1\sim(1,2,1)$] and uses $S_2\sim(1,2,3)$ and $\f\sim(1,2,-1)$.}\label{fig:KNT_diff_f}
\end{figure}

In the preceding section we classified the models that give mass via Figure~\ref{fig:general_KNT}, all of which utilized a Majorana beyond-SM fermion, $\f$. Recall that the one-loop model of Ma~\cite{Ma:2006km} and its triplet variant~\cite{Ma:2008cu} both employ Majorana fermions. However, these models belong to a generalized class which includes models with Dirac fermions~\cite{Law:2013saa}. One might expect variants of the three-loop models to exist which similarly employ  Dirac fermions. We shall see that this is the case. However, to allow for Dirac fermions while retaining a three-loop diagram with the same topology as Figure~\ref{fig:general_KNT}, one must allow for different SM fermions in the left- and right-loops. In this section we therefore relax the demand that the SM fermions in the left- and right-loops are the same. This allows for models with Dirac fermions, as indicated by Figure~\ref{fig:KNT_diff_f}, which shows the general three-loop diagram in this case. In the figure, both $f$ and $f'$ are SM fermions, with $f\ne f'$ assumed. The fields in the inner loop are taken odd under a $Z_2$ symmetry, with all other fields being even.

We performed a systematic search for viable models in this case. These models turn out to be more complex than the simpler case with a Majorana  fermion, requiring additional beyond-SM scalars in addition to the Dirac fermion $\f=\f_L+\f_R$. For concreteness we discuss the case with $f^c_{L,R}=e^c_{L,R}$, while $f'\ne f$ is assumed arbitrary. We considered fermion multiplets $\f$ no larger than the adjoint representation and focused on color-less $\f$, which is generally required to allow a suitable DM candidate. The models with Dirac mediators are of course free of gauge anomalies. We find that for $f'_{R,L}=u^c_{L,R}$ no viable models arise and  similarly with $f'_{L,R}=u_{L,R}$.

A number of models that achieve neutrino mass and contain DM candidates arise if one employs down-type quarks. As an illustrative example, consider the case with $f'_{L,R}=d^c_{R,L}$, which gives the new  models listed in Table~\ref{KNT_diff_f}. In all of the models, one uses a Dirac beyond-SM fermion and the Lagrangian generically contains the terms
\bea
\mathcal{L}&\supset& \overline{L^c}\,L\,S_1+\overline{F_R}\,e_R^c\,\phi_1+\overline{Q^c}\,F_L\,\phi_2+\overline{L^c}\,d_R^cS_2 - S_1^\dagger\phi_1^\dagger\phi_2^\dagger S_2^\dagger,
\eea
where we suppress coupling constants for simplicity. These terms are sufficient to break lepton number symmetry. Here $S_1\sim(1,1,2)$ and $S_2\sim(3,2,1/3)$ are common to all the models, while the quantum numbers for $\f$ and $\phi_{1,2}$ vary in accordance with Table~\ref{KNT_diff_f}.

In all cases the DM is an inert N-tuplet scalar; there are cases with singlet, doublet and triplet DM, depending on the quantum numbers for the Dirac fermion. We expect the inert-triplet DM to have mass $M_\dm\sim2$~TeV, in line with previous studies~\cite{Araki:2011hm}. The inert doublet DM is largely constrained to the heavier region of viable parameter-space with $M_\dm\gtrsim 500$~GeV, which is beyond the reach of the LHC~\cite{Goudelis:2013uca}   but can be within reach of  direct-detection experiments~\cite{Klasen:2013btp}.


\begin{table}
\centering
\begin{tabular}{|c|c|c|c|c|}\hline
\ \ Model\ \ &
$\mathcal{F}$ & $\phi_1$&$\phi_2$&Dark Matter \\
\hline
$(A)$& $\ \ (1,1,2)\ \ $ &$\ \ (1,1,0)\ \ $&\ \ $(\bar{3},2,-7/3)$\ \ & Inert Singlet\\ 
\hline
$(B)$& $\ \ (1,2,1)\ \ $ &$\ \ (1,2,-1)\ \ $&\ \ $(\bar{3},1,-4/3)$\ \ & Inert Doublet\\ 
\hline
$(C)$& $\ \ (1,2,1)\ \ $ &$\ \ (1,2,-1)\ \ $&\ \ $(\bar{3},3,-4/3)$\ \ & Inert Doublet\\ 
\hline
$(D)$& $\ \ (1,3,2)\ \ $ &$\ \ (1,3,0)\ \ $&\ \ $(\bar{3},2,-7/3)$\ \ & Inert Triplet\\ 
\hline
$(E)$& $\ \ (1,2,3)\ \ $ &$\ \ (1,2,1)\ \ $&\ \ $(\bar{3},1,-10/3)$\ \ & Inert Doublet\\ 
\hline
$(F)$& $\ \ (1,2,3)\ \ $ &$\ \ (1,2,1)\ \ $&\ \ $(\bar{3},3,-10/3)$\ \ & Inert Doublet\\ 
\hline
\end{tabular}
\caption{\label{KNT_diff_f} Models with radiative neutrino mass via Figure~\ref{fig:KNT_diff_f} with DM propagating in the inner loop. In all cases we use $f^c_{L,R}=e^c_{L,R}$ as SM leptons,  $f'_{L,R}=d^c_{R,L}$ as down-type SM quarks, and the beyond-SM scalars $S_1\sim(1,1,2)$ and $S_2\sim(3,2,1/3)$.}
\end{table}

For some models in Table~\ref{KNT_diff_f} the fermion $\f$ contains a neutral component. However, these fermions are not viable DM candidates as they have nonzero hypercharge and therefore have  tree-level electroweak interactions with detectors. The neutral components are generally split by radiative corrections but the splitting is tiny, being proportional to the SM neutrino mass scale. Thus, although the neutral fermions are technically pseudo-Dirac particles, for all practical purposes they behave like Dirac particles and are therefore excluded by direct-detection constraints. Consequently only scalar DM is possible in these models.

In the alternative case where $f'_{L,R}=d_{L,R}$ while $f_{L,R}=e_{L,R}$ is retained  we do not find any viable models. We don't consider the case with Majorana fermions in Figure~\ref{fig:KNT_diff_f}; even if a viable model could be found it would also give mass via the diagram in Figure~\ref{fig:general_KNT}, making any contribution from Figure~\ref{fig:KNT_diff_f} redundant.

\section{A Related Class of Three-Loop Models with Dark Matter\label{sec:general_3loop_withDM}}

\begin{figure}[ttt]
\begin{center}
        \includegraphics[width = 0.6\textwidth]{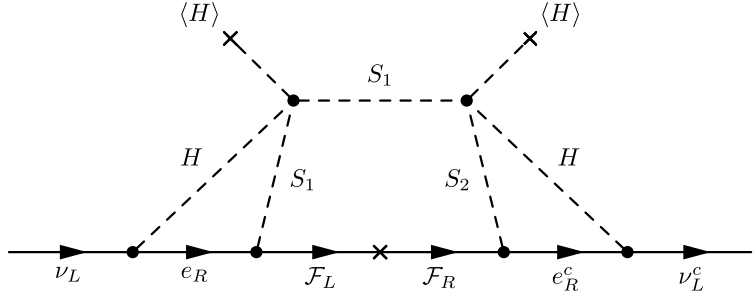}
\end{center}
\caption{A three-loop diagram for radiative neutrino mass, where $S_{1,2}$ and $\mathcal{F}$ are beyond-SM fields and DM propagates in the inner loop. The simplest case has inert-doublet DM [$S_1\sim(1,2,1)$] and uses $S_2\sim(1,2,3)$ and $\f\sim(1,2,-1)$.}\label{fig:related_3loop}
\end{figure}

In the preceding we first identified a class of models that included the KNT model, all members of which contained a Majorana beyond-SM fermion. We then allowed a mixture of quarks and electrons to propagate in the loop diagram, arriving at related models (with the same three-loop topology) that use Dirac fermions. In all of these models, the three-loop diagram contains two mass-insertions on the internal SM-fermion lines. These supply the Higgs VEVs which allow one to write the mass as $m_\nu\propto \langle H\rangle^2/\Lambda$ for some effective new-physics scale $\Lambda$. Another way to generalize the KNT model, while retaining three-loop neutrino masses with DM in the inner loop, is to modify the topology of the diagram. In this section we consider related models which have two insertions of the Higgs VEV on the scalar lines, rather than the fermion lines. This allows new models  that  employ a Dirac fermion $\f$, one of which is simpler than the models described in Table~\ref{KNT_diff_f}.

We draw the Feynman diagram for the most-promising model of this type in Figure~\ref{fig:related_3loop}. In general one can replace $e_R$ with any right-chiral SM fermion (with $H$ also replaced by a beyond-SM multiplet), and let the two occurrences of $S_1$ become distinct fields. However, in the case where $e_R$ is replaced with $e_L^c$, we find that the models also give one-loop masses that are expected to dominate the three-loop mass. For example, replacing $H\rightarrow S\sim(1,1,2)$ and using $\f=\f_L+\f_R\sim(1,1,2)$, $S_1\sim(1,2,1)$ and $S_3\sim(1,2,-3)$, one obtains a variant of Figure~\ref{fig:related_3loop}, but the model also allows the one-loop diagram from Refs.~\cite{Aoki:2011yk,Law:2013saa}. We therefore disregard the case with $e_R\rightarrow e_L^c$.

\begin{figure}[ttt]
\begin{center}
        \includegraphics[width = 0.6\textwidth]{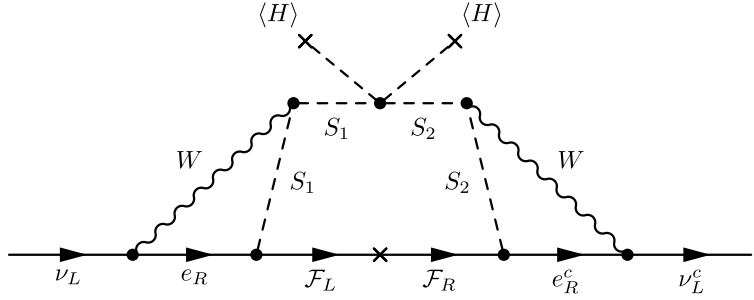}
\end{center}
\caption{A three-loop diagram for radiative neutrino mass for the model shown in Figure~\ref{fig:related_3loop}, with $S_1\sim(1,2,1)$, $S_2\sim(1,2,3)$ and $\f\sim(1,2,-1)$. The diagram persists in unitary gauge. }\label{fig:unitary_3loop}
\end{figure}

A systematic study of the case shown in Figure~\ref{fig:related_3loop}, with internal $e_R$, reveals a single viable model. The SM is extended to include $\f=\f_L+\f_R\sim(1,2,-1)$ and two scalars, $S_1\sim(1,2,1)$ and $S_2\sim(1,2,3)$. The fields in the inner loop are taken to be odd under the $Z_2$ symmetry, $\{S_{1,2},\,\f\}\rightarrow\{-S_{1,2},\,-\f\}$, so the Lagrangian contains the new terms
\bea
\mathcal{L}&\supset& \overline{\f_L}\,e_R\, S_1 \,+\,\overline{e_R^c}\,\f_R\,S_2 \;-\;\overline{\f_L} \,M_\f\, \f_R\;-\;  (S^\dagger_1\,H)^2\, -\,S^\dagger_2 H\tilde{H}^\dagger S_1+\mathrm{H.c.},\label{eq:eR_related_model}
\eea
where we suppress coupling constants. These terms are sufficient to break lepton number symmetry and generate Figure~\ref{fig:related_3loop}. The DM is a neutral component of $S_1\sim(1,2,1)$, and is therefore an inert doublet.  Note that this model requires less multiplets than the models in Table~\ref{KNT_diff_f}. One notices that the diagram in Figure~\ref{fig:related_3loop} vanishes in unitary gauge, however, there is a related diagram involving $W$ bosons that persists, successfully generating neutrino mass at the three-loop level, as shown in Figure~\ref{fig:unitary_3loop}. The form in Figure~\ref{fig:related_3loop} is easier to generalize for cases with SM quarks and non-SM scalars inside the left- and right-loops, which is why we display it.

\begin{figure}[ttt]
\begin{center}
        \includegraphics[width = 0.6\textwidth]{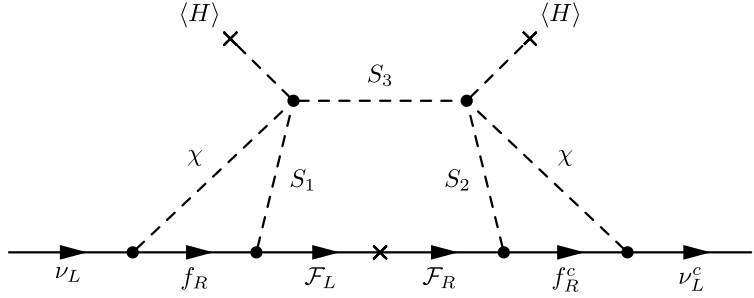}
\end{center}
\caption{A three-loop diagram for radiative neutrino mass with inert-doublet DM. Quantum numbers for the various multiplets are  listed in Table~\ref{related_3loop_d}. }\label{fig:related_3loop_dR}
\end{figure}

One can also use colored fields in place of $e_R$. It seems that viable models can be found, though the particle content becomes more involved. The general loop diagram appears in Figure~\ref{fig:related_3loop_dR}. As an example, for the case with  $f_R= d_R$ one requires the multiplets $\chi\sim(3,2,1/3)$, $S_1\sim(\bar{3},2,5/3)$, $S_2\sim(\bar{3},2,-1/3)$,  and $S_3\sim(1,2,1)$, along with the fermion $\f\sim(1,2,1)$. The multiplets $\{\f,\,S_{1,2,3}\}$ are odd under the $Z_2$ symmetry, so the Lagrangian contains
\bea
\mathcal{L}&\supset& \overline{d_R}\,L\,\chi+\overline{\f_L}\,d_R\, S_1 \,+\,\overline{d_R^c}\,\f_R\,S_2 \;-\;\overline{\f_L} \,M_\f\, \f_R\;-\; \chi^\dagger S_1^\dagger \tilde{H}^\dagger S_3- \chi^\dagger S_2^\dagger S_3^\dagger H  +\mathrm{H.c.}.\nonumber
\eea
Lepton number symmetry is again broken and three-loop neutrino masses result. Retaining the replacement $e_R\rightarrow d_R$ in Figure~\ref{fig:related_3loop}, there are a number of other combinations of the multiplets $\f$, $\chi$ and $S_{1,2,3}$ that allow the three-loop diagram. However, the case just mentioned appears to be the only combination that gives a viable DM candidate. All other combinations either lack non-colored neutral scalars or contain a neutral field with nonzero hypercharge that is ruled out by direct-detection constraints. Viable models with $f_R=d_L^c$ can also be found. There are some additional options in this case as one can employ either singlet or triplet $SU(2)$ scalars. Similarly for $f_R=u_R$ and $f_R=u_L^c$, though the latter case has overlap with the case of $f_R=d_L^c$ and can give multiple diagrams.

We summarize the particle content for the viable models with internal quarks in Table~\ref{related_3loop_d}. One observes that the number of new multiplets required is somewhat larger than the case with $f_R=e_R$ and $\chi\rightarrow H$ in Figure~\ref{fig:related_3loop} . The implementation with colorless beyond-SM fields  is clearly  simpler and appears to be the favored case for this loop topology.

Before concluding let us mention a couple of important constraints. In general, flavor-changing constraints restrict the viable parameter space for three-loop models of neutrino mass. The size of the effect is dependent on the details of the $Z_2$-odd sector and is therefore model-dependent.  For example, one-loop $\mu\rightarrow e+\gamma$ decays give important constraints in the KNT model~\cite{Ahriche:2013zwa} and the triplet variant~\cite{Ahriche:2014cda}, though viable parameter space exists in each case. Similar effects are expected for the model with non-colored multiplets in Figure~\ref{fig:unitary_3loop}. The models with new colored fields generate additional flavor-changing diagrams involving colored internal fields.\footnote{Related colored contributions to $\mu\rightarrow e+\gamma$ appear in the colored variant of the Zee-Babu model~\cite{Kohda:2012sr}.} The severity of the constraints will depend on the given field content and the requisite  DM mass for the given model (e.g., for $\f_R\sim(1,3,0)$ DM one has $M_\dm\sim2$~TeV while for $\f_R\sim(1,1,0)$ one expects $M_\dm=\mathcal{O}(100)$~GeV). A detailed study is required to determine the viable parameter space for the individual models. We also note that some models with colored beyond-SM multiplets can cause the proton to decay. However,  a baryon number symmetry can be imposed in all such cases to ensure proton longevity without disturbing the neutrino masses.

\begin{table}
\centering
\begin{tabular}{|c|c|c|c|c|c|}\hline
\ \ $f_{R}$\ \ &
$\mathcal{F}$ & $S_1$&$S_2$&$S_3$&$\chi$ \\
\hline
$d_{R}$& $\ \ (1,2,1)\ \ $ &$\ \ (\bar{3},2,5/3)\ \ $&\ \ $(\bar{3},2,-1/3)$\ \ & $(1,2,1)$\ &\ $(3,2,1/3)$\\ 
\hline
$d_{L}^c$& $\ \ (1,2,1)\ \ $ &$\ \ (3,1\oplus3,4/3)\ \ $&\ \ $(\bar{3},1\oplus3,-2/3)$\ \ & $(1,2,1)$\ & \ $(\bar{3},1\oplus3,2/3)$\\ 
\hline
$u_{R}$& $\ \ (1,2,1)\ \ $ &$\ \ (\bar{3},2,-1/3)\ \ $&\ \ $(\bar{3},2,-7/3)$\ \ & $(1,2,1)$\ &\ $(3,2,7/3)$\\ 
\hline
$u_{L}^c$& $\ \ (1,2,1)\ \ $ &$\ \ (3,3,4/3)\ \ $&\ \ $(3,3,-2/3)$\ \ & $(1,2,1)$\ & \ $(\bar{3},3,2/3)$\\ 
\hline
\end{tabular}
\caption{\label{related_3loop_d} Models with radiative neutrino mass via Figure~\ref{fig:related_3loop_dR} with DM propagating in the inner loop and internal SM quarks. In all cases the DM resides in the inert doublet $S_3$.}
\end{table}


\section{Conclusion\label{sec:conclusion}}

We studied a class of models with three-loop  neutrino masses that depend on the existence of DM. The models contain a $Z_2$-odd sector that is sequestered from SM neutrinos and propagates in the inner-loop of the mass diagram. The simplest models have Majorana DM and include the proposal of KNT; we identified the related variants of this model and, in particular,  presented the viable colored variants. By extending the particle content and/or modifying the loop topology, we found additional related models  that use a Dirac mediator and give inert scalar DM, with singlet, doublet and triplet cases possible.  The simplest such model generates neutrino mass via the diagram shown in  Figure~\ref{fig:unitary_3loop}. This model appears to be the favoured ``related model" and  is worthy of further study. We shall study the new colored variants and the simple model of Figure~\ref{fig:unitary_3loop} in future works.

\section*{Acknowledgments\label{sec:ackn}}
CSC is supported by the National Center for Theoretical Sciences, Taiwan with grant
number NSC 103-2119-M-007-003. KM is supported by the Australian Research Council.
\appendix


\end{document}